# Relaxations and Relaxor-Ferroelectric-like Response of Nanotubularly Confined Poly(vinylidene fluoride)

Jaime Martín[1,]*, Amaia Iturrospe[2], Andrea Cavallaro[1], Arantxa Arbe[2], Natalie Stingelin[1], Tiberio A. Ezquerra[3], Carmen Mijangos[4], Aurora Nogales[3,]*

[1] Department of Materials, Imperial College London, Exhibition Road, London, SW7 2AZ, UK

[2] Centro de Física de Materiales (CFM) (CSIC−UPV/EHU), Materials Physics Center (MPC), Paseo Manuel de Lardizabal 5, 20018 San Sebastián, Spain

[3] Instituto de Estructura de la Materia IEM-CSIC, C/ Serrano 121, Madrid 28006, Spain

[4] Instituto de Ciencia y Tecnología de Polímeros ICTP-CSIC, C/ Juan de la Cierva 3, Madrid 28006, Spain

**Abstract**

Herein, we elucidate the impact of tubular confinement on the structure and relaxation behaviour of poly(vinylidene difluoride) (PVDF) and how these affect the para-/ferroelectric behavior of this polymer. We use PVDF nanotubes that were solidified in anodic aluminum oxide (AAO) templates. Dielectric spectroscopy measurements evidence a bimodal relaxation process for PVDF nanotubes: besides the bulk-like α-relaxation, we detect a notably slower relaxation that is associated with the PVDF regions of restricted dynamics at the interface with the AAO pore. Strickingly, both the bulk-like and the interfacial relaxation tend to become temperature independent as the temperature increases - a behavior that has been observed before in inorganic relaxor ferroelectrics. In line with this, we observe that the real part of the dielectric permittivity of the PVDF nanotubes exhibits a broad maximum when plotted against the temperature, which is, again, a typical feature of relaxor ferroelectrics. As such, we propose that in nanotubular PVDF, ferroelectric-like nanodomains are formed in the amorphous phase regions adjacent to the AAO interface. These ferroelectric nanodomains may result from an anisotropic chain conformation and a preferred orientation of local dipoles due to selective H-bond formation between the PVDF macromolecues and the AAO walls. Such relaxor-ferroelectric-like behaviour has not been observed for non-irradiated PVDF homopolymer; our findings thus may enable in the future alternative applications for this bulk commodity plastic, e.g., for the production of electrocaloric devices for solid-state refrigeration which benefit from a relaxor-ferroelectric-like response.

**Introduction**

Understanding how the properties of macromolecular materials confined to the nanoscale differ from those of their bulk counterparts is not only of fundamental interest for polymer scientists but is becoming technologically increasingly important because applications based on, e.g., printable optoelectronics, biomedicine and photonics, more and more frequently employ sub-micron polymeric architectures in specific elements, devices and more integrated structures [1-5]. As a consequence, it is essential that we gain insights into how spatial confinement affects the properties of high polymers in order to establish the means to manipulate and control from the outset their optoelectronic [6], mechanical [7], thermal [8], and ferro-/piezoelectric features when used in nanostructures.

The reason why typical attributes of polymers are affected – and thus can be influenced – by spatial confinement, is the fact that spatial confinement alters the structural and dynamical behavior of the individual macromolecules and impacts the processes that exhibit characteristic length scales in the nanometer range. Examples of such processes include among onthers: melting [9], nucleation [10], crystal growth [11], , segmental dynamics [12-13], chain dynamics [14], and reptation through entanglements.[15]

When describing confinement of macromolecular matter, two fundamental aspects of spatial restriction must be considered: The first is the degree of spatial constrain, which depends on the available space and the characteristic length scales of the specific dynamical or structural process. The second is the geometry of spatial constrain, which is determined by the shape of the space where the dynamical or structural process occurs and, thus, dictates whether the process proceeds in a 1D-, 2D- or 3D fashion. Confinement occurs, for instance, in one dimension in thin films or block copolymers that phase separate into lamellae; in two dimensions in nanowires or cylindrical phases of block copolymers; and in three dimensions in nanoparticles or block copolymers that form spherical phases – all leading to different properties even when the various structures are made of the same material.

Among these simple confinement geometries, nanotubularly confined structures are attractive systems to elucidate various geometrical aspects of nanoscale confinement in polymers due to their versatility and the simplicity with which they can be produced. Moreover, hollow tubes can be produced, allowing additional functions, resulting from the empty inner cavity, to be introduced [16]. In general, such polymeric hollow nanotubes can be considered to be systems between one- and two-dimensional confinement. On the one hand, the wall of a typical polymer nanotube produced in, e.g., an anodic aluminium oxide (AAO) nanoporous templates, is of a thickness within the tens-of-nanometer-range[17], which leads to the confinement of the individual macromolecules along the wall's radial direction. On the other hand, such nanotubes can frequently be micrometer long; hence, there is no confinement along this direction. The confinement along the azimuthal direction needs also to be taken into account. It depends on both the radius of curvature of the nanotube (and, thus, its radius), and the abitility of the the structural or dynamical process in question to accommodate to this curvature.

Various routes towards polymeric hollow nanotubes have been explored, including rolling thin films up [12], self-assembly of block copolymers [13], use of cyclic polymers [14] and polypeptoids [15], as well as templating, using both hard [16] and soft [17] templates. We selected to use infiltration of AAO nanoporous templates [18] — a method that enables fabrication of hollow nanotubes of any macromolecular matter that can be processed from solution or the melt [11, 19]. Molten PVDF can be infiltrated within these nanopores and solidified in confinement, which give rise to the formation of PVDF nanotubes. The AAO templates are rigid and their cylindrical pores are highly homogeneous and of low size dispersion. The AAO is inert and stable below 500 °C, i.e. in the temperature range where most of the relevant physical processes (e.g., nucleation, melting) in polymers occur. This enables elucidation of the dynamical and structural features under such nanotubular confinement.

PVDF, that is (-[$CH_2$-$CF_2$]$_n$-), was chosen as it is a model semicrystalline polymer allowing us to analyse how nanotubular confinement affects the internal microstructure of polymers both with respect to their amorphous and crystalline phases. Secondly, and more importantly for our objectives here, PVDF features a high transverse dipole moment that facilitates the assessment of its dynamics by dielectric spectroscopy (DS) and allows comparison with structural information obtained from differential scanning calorimetry (DSC), wide-angle X-ray scattering (WAXS) and small angle X-ray scattering (SAXS). Finally, PVDF can feature ferro- and piezoelectric features which assists to gain additional information on any conformational changes that are introduced in confinement when compared to bulk PVDF processed under similar conditions.

**Experimental Section**

*Anodic Aluminum Oxide.* Self-ordered anodic aluminum oxide (AAO) with an average pore diameter of 400 nm, a pore depth of 100 μm was prepared by two-step anodization of aluminum using phosphoric acid as electrolyte at 205 V following procedures described elsewhere [18]. A representative scanning electron microscopy (SEM) top view is shown in Figure 1a, where the honeycomb porous structure can be observed. The lattice constant of the hexagonal cell is 480 nm.

*Preparation of poly(vinylidene fluoride) nanotubes.* Commercially available poly(vinylidene fluoride) (PVDF) (Aldrich, Ltg. $M_w$=180,000 g/mol, $M_n$=71,000 g/mol) was used. A PVDF film was placed on the surface of the AAO templates at 260 ºC for 15 minutes under nitrogen atmosphere. Under these conditions, the molten PVDF wets the AAO templates in the complete wetting regime [19-20]. In that wetting scenario, liquid precursor films with thicknesses of tens of nm spread over nanopore walls. In order to avoid the excessive thickening of the precursor films during the infiltration [21], a relatively short annealing time was used (15 min). The samples were then rapidly cooled down to room temperature, which provoked the solidification of the tubular precursor films and, thus, the formation of the PVDF nanotubes. The residual PVDF films located on top of the AAO templates was then removed with a sharp blade so that the nanotubes within the AAO were isolated entities separated from each other. After that, the samples were again annealed to 260 ºC for 5 min. The samples used for the microstructural characterization were then cooled at 1 ºC/min in order to favor crystallization, whereas the samples used for the dynamical study were rapidly quenched to 25 ºC immersing them in water.

*Scanning electron microscopy*: The morphological characterization of the samples was conducted by scanning electron microscopy (SEM). The nanotubes were released from the template employing NaOH (10 wt.%).

*Differential Scanning Calorimetry.* A differential scanning calorimeter (DSC) (Perkin-Elmer DSC-7) was used for the thermal characterization of the samples. Heating runs at 10 ºC/min were carried out under a constant flow of nitrogen. For the evaluation of the crystallinity, the 1$^{st}$ heating runs were studied, as these reflect the melting of crystals formed during the processing of the nanotubes. For the DSC study, the aluminum substrates attached to the AAO templates were selectively etched employing a mixture of 1.7 g $CuCl_2 \cdot H_2O$, 50 ml concentrated HCl and 50 ml deionized water.

*Wide angle X-ray scattering.* The 2θ scan analysis of the sample was performed by X-ray diffraction (XRD) in a four circle goniometer Panalytical Empyrean; CuK$_\alpha$ radiation (λ = 1.54 Å) in line focus. The XRD poles figure analysis (020), (110) and (021) have instead been recorded in a Panalytical X'-Pert CuKα radiation in point focus.

*Small angle X-ray scattering.* Small-angle X-ray scattering (SAXS) was used to probe the supracrystalline structure of our samples. Experiments were conducted on a Rigaku 3-pinhole PSAXS-L equipment operating at 45 kV and 0.88 mA. The MicroMax-002+ X-ray generator system is composed by a microfocus sealed tube source module and an integrated CuK$_\alpha$ X-ray generator unit. The flight path and the sample chamber in this equipment are under vacuum. The scattered X-rays are detected on a two-dimensional multiwire X-ray detector (Gabriel design, 2D-200X). This gas-filled proportional type detector offers a 200 mm diameter active area with ca. 200 μm resolution. The azimuthally averaged scattered intensities were obtained as a function of scattering vector $q$, $q = 4\pi\lambda^{-1} \sin(\theta)$. Reciprocal space calibration was done using silver behenate as standard. Samples were placed in transmission geometry, with sample-to-detector distances between 0.5 and 2 m. For the SAXS experiments, the nanotubes were released from the template with a NaOH solution (10 wt.%), thoroughly rinsed with water and placed between mica slides.

*Dielectric Spectroscopy.* Dielectric spectroscopy (DS) measures the complex dielectric permittivity ε* = ε´ - iε´´ as a function of frequency, where ε´ is the dielectric constant and ε´´ is the dielectric loss. PVDF nanotubes embedded into the AAO templates were placed between two electrodes of 1 cm diameter. Nanotubes laid thus normal to electrodes. The dielectric spectroscopy measurements were performed over a broad frequency window, $10^{-1} < F(Hz) < 10^7$ by means of a Novocontrol system integrating a dielectric interface (ALPHA) and a temperature control by nitrogen jet (QUATRO) with a temperature error, during every single sweep in frequency, of 0.1 K. The analysis of the results was performed by means of the WinFit program (Novocontrol).

**Results and discussion**

**Nanotube fabrication**

We produced hollow nanotubes by melt infiltration of PVDF at 260 ºC into nanoporous AAO templates (pore diameter = 400 nm, pore length = 100 μm) followed by cooling to room temperature at 1 ºC /min, according to procedures first reported by Steinhart et al. [22-23]. A scanning electron micrograph of an AAO template is presented in Fig. 1a. Fig. 1b shows a low-magnification scanning electron micrograph of a mat of nanotubes that were produced in this manner, after they had been released from the porous template. The hollow central cavity of the nanotubes can be clearly distinguished in Fig. 1c. The nanotubes have a length of 100 μm and an outer diameter of 400 nm, while the thickness of the tube walls is around 35 nm consistent with the observations made in Ref. [22].

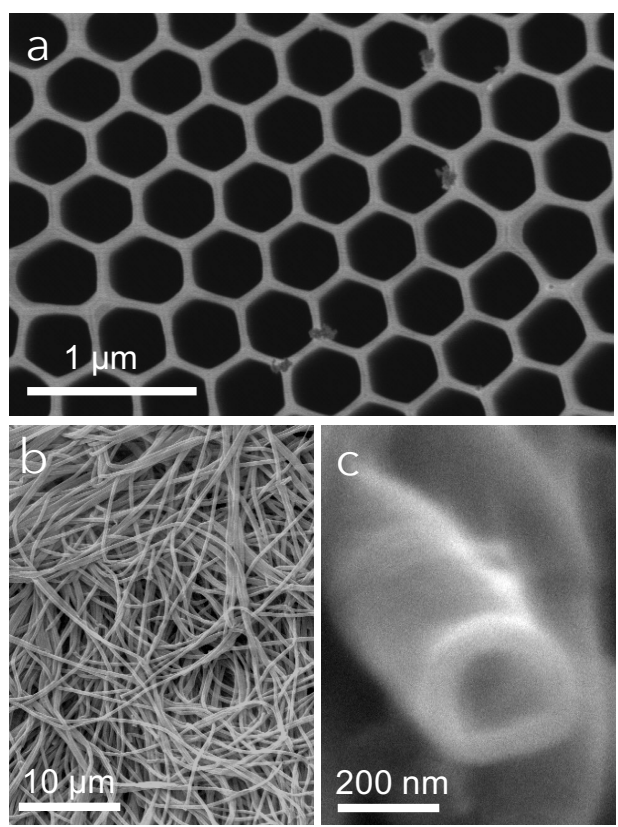

**Figure 1**. SEM micrographs of an AAO template used in this study (a) and PVDF nanotubes after being extracted from the AAO template (b, c). From (c) it is clear that hollow nanotubes are produced in agreement with previous reports by Steinhart et al. [29] .

**Effect of nanotubular confinement on the microstructure**

Since it is well known that confinement can strongly impact the structural features of polymers [22-23] and generally leads to a lower degree of crystallinity than bulk solidification [24], we first analyzed the melting enthalpies $\Delta H_f$ of our PVDF nanotubes deduced from the first DSC scans (see Fig. 2) and compared these with values obtained for bulk crystallized material. Note that first heating scans have been considered as these account for the microstructure develop during the nanoprocessing of our PVDF nanotubes. For this analysis we calculated the mass of PVDF contained in the nanotubes utilizing geometrical considerations and assuming a single honeycomb domain for the whole template, as outlined in the Supporting Information Fig S1. We obtain $\Delta H_{f,\ nanotube}$, ≈ 37 J/g, which corresponds to a degree of crystallinity $X_{c,\ nanotube}$, ≈ 38 % assuming that $\Delta H_f$ for a 100% crystalline material is 105 J/g [25]. This value is only slightly lower (by 6 %) than that of bulk PVDF crystallized under the same conditions ($X_{c,\ bulk}$ ≈ 44%; $\Delta H_{f,\ bulk}$ ≈ 46 J/g) and agrees with previously determined values by Shigne et al., although they employed a different method to deduce the mass of the nanotubes within the AAO templates [26].

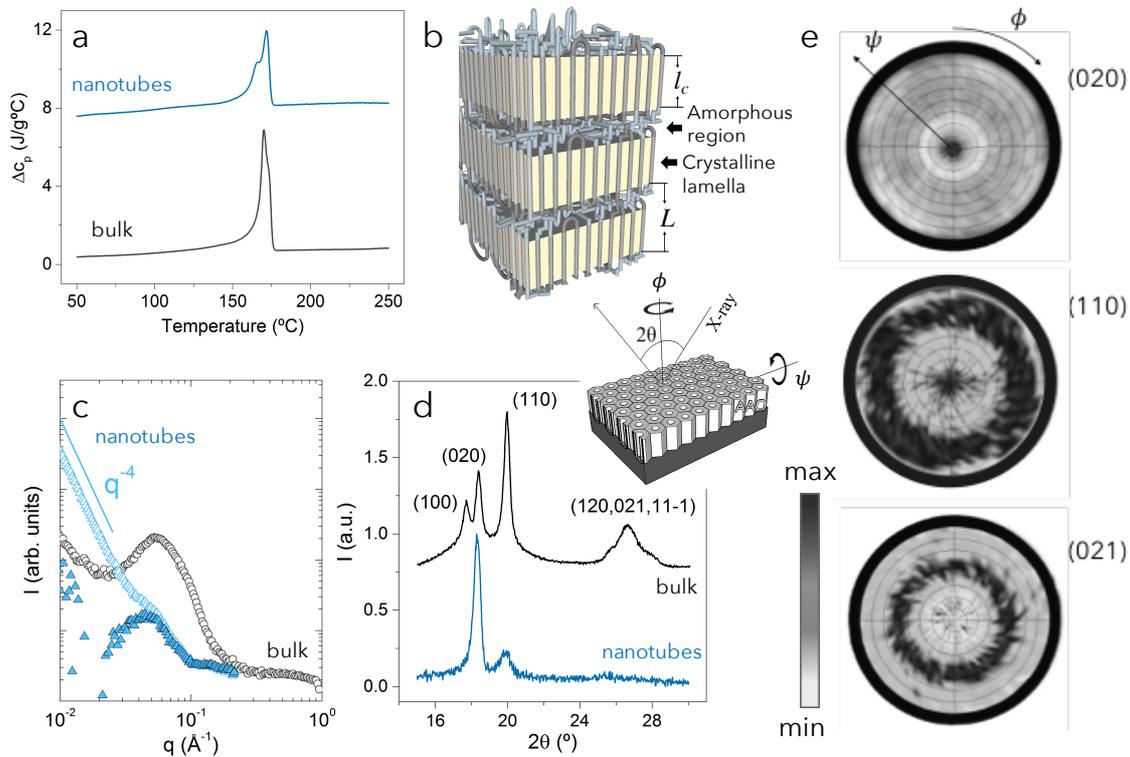

**Figure 2**. (a) Representative DSC first heating thermograms recorded at 10ºC/min of PVDF hollow nanotubes (blue curve) and bulk PVDF (grey curve) (the $c_p$ values are scaled for clarity), both solidified from the melt by cooled to room temperature at 1ºC/min. (b) Schematic of the stack of polymeric lamellar crystals, where the crystalline lamellar thickness, $l_c$, and the long period, $L$, are identified. (c) SAXS intensity patterns measured for bulk-crystallized PVDF (grey circles) and PVDF nanotubes released from their AAO template (open blue triangles). Blue line is a guide for the eyes indicating a slope of -4 in this double logarithm scale. Solid blue triangles show the SAXS contribution from the PVDF after subtracting the $q^{-4}$ dependence. (d) WAXS patterns of bulk PVDF (black like) and (d) PVDF nanotubes (blue line) for $q$ parallel to the long axis of the nanotubes, i.e. $\psi=0°$. The geometry of the experiments conducted and the scanned angles are indicated the schematic given in the inset in (d). (e) XRD pole figures of the (020), (110), and (021) reflections measured in PVDF nanotubes. Each circular line along the radial direction in the pole figures corresponds to $\psi = 10°$.

The only moderate reduction of the degree of crystallinity of PVDF solidified in AAO templates compared to the one of bulk material must be associated with the fact that the crystal growth process is not strongly hindered, similar to what was observed for hollow nanotubes of PVDF-based copolymers and their bulk structures [27]. This implies that once crystallites are nucleated within the nanotube wall, they are allowed to propagate along the longitudinal as well as the azimuthal direction of the nanotubes. Note that the reported degree of crystallinity of PVDF nanorods with a diameter of 35 nm [12, 22] is significantly lower because stronger confinement effects that can take place in these structures compared to our (hollow) PVDF nanotubes.

Other valuable information can be obtained from the DSC data presented in Fig. 2. It is, for instance, striking that we do observe essentially identical melting temperatures ($T_m$) for the PVDF nanotubes ($T_{m\ peak,\ nanotube}$ = 171 ºC) and the bulk-crystallized material ($T_{m\ peak,bulk}$ = 170 ºC). This implies, according to the Gibbs-Thomson theory [28], that the thicknesses of the crystalline lamellae $l_c$ (see Fig 2b) in these semicrystalline structures are similar – if not equal. Hence, assuming a typical semicrystalline microstructure of polymers that is comprised of crystalline lamellae stacks and interlamellar amorphous regions [29], our DSC results (slight change in degree of crystallinity, almost negligible change in metling temperature) can be explained either with the existence of an independent, additional amorphous region in the PVDF nanostructures like those demonstrated elsewhere [30] or assuming that the interlamellar amorphous regions are somewhat expanded in the confined PVDF structures. Such an expansion of the amorphous regions without changing the thickness of the crystalline moieties would lead in a change of the so-called long period $L$ [31] — which is the periodicity given by a stack comprised of an ordered lamellae and an unordered regions (see Fig. 2b). Since $L$ can be deduced from scattering experiments, SAXS measurements were achieved in order to prove wether the reduction of the crystallinity was due to an independent amorphous region or due to an enlargement od the lamellar stack. Figure 2b shows the SAXS intensity for bulk PVDF as well as for our PVDF nanotubes after they had been released from the AAO templates. The SAXS pattern for the bulk PVDF is characterized by a maximum at $q \sim 0.052$ Å$^{-1}$, corresponding to a characteristic long period, $L_{bulk} \approx 12.1$ nm. The intensity scattered by the PVDF nanotubes exhibits a strong contribution at low $q$ that decreases with a $q^{-4}$ dependence as well as a peak-like contribution that seems to be superimposed to the former feature. The $q^{-4}$ dependence can be attributed to the Porod region caused by the scattering of long cylindrical structures (here: the PVDF tubes). After subtracting this scattering contribution, a peak can be indentified in the SAXS pattern of the PVDF nanotubes, which originates from the presence of periodic stacks of crystalline lamellae/amorphous regions in the PVDF tubes. The maximum scattering of this feature is located at $q \sim 0.044$ Å$^{-1}$, from which we deduce $L_{nanotube} \approx 14.2$ nm; i.e. somewhat larger than $L_{bulk}$.

In the continuous lamellar stack model $L$ can be expressed as $l_c = L \cdot X_c$, where $l_c$ is the lamellar thickness mentioned above. Accordingly, we can deduce $l_{c,bulk} \approx l_{c,nanotube} \approx 5.3$ nm using the obtained $L$-values for the two PVDF systems from SAXS and the $X_c$-values from DSC, and assuming that both PVDF samples feature the same crystal structures. The finding that $l_{c,bulk} \approx l_{c,nanotube}$ supports the hypothesis that the nearly identical temperatures observed in DSC measurements for both bulk PVDF and PVDF nanotubes are due to similar lamellar crystal thickness.

Let us analyze our data in a different way. If $l_{c,bulk}$ is assumed to be equal to $l_{c,nanotube}$ because both materials crystallized in the same crystalline phase (Fig. 2d) and show equal melting temperatures (Fig. 2a), $X_c$ can be deduced from SAXS, as $X_c = l_c / L,$ if the continuous lamellar stack model is assumed. The $X_{c,nanotube}$ thus obtained amounted to 37 %, which is very similar to that obtained via the enthalpy of fusion ($X_{c,nanotube}= 38\%$). Therefore, this result seems to indicate that the continuous lamellar stack model is valid for our nanotubes and thus that the great majority of the amorphous PVDF chains in nanotubes are contained between crystalline lamellae. Hence, our results suggest the absence of another, distinct amorphous phase that potentially could be present in slowly crystallized PVDF nanotubes at the PVDF-AAO interface and/or the PVDF-air interfaces, such as those demonstrated by Sonnenberger et al. [30]. We should point out, though, that the calculation of the $L_{nanotube}$ value by SAXS is subject to a moderate uncertainty, which can alter the result of the crystallinity.

In order to scrutinize whether the bulk crystallized samples and the nanotubes feature a similar crystalline arrangement and to analyze potential preferred orientation of the crystalline moieties induced by the solidification in confined state, we went on to perform WAXS measurements. Fig. 2d shows the $2\theta$-WAXS scans for bulk PVDF and PVDF nanotubes, still embedded in the AAO template. The scattering vecor, $q$, was thereby kept parallel to nanopore long axis (see schematic in Fig. 2d).

In the $2\theta$-range of 15º to 30º investigated here, the bulk PVDF pattern shows the characteristic reflections of the orthorhombic unit cell of the α-phase with the following dimensions: $a$ = 4.96 Å, $b$ = 9.64 Å and $c$ = 4.62 Å[32-33] (a schematic of the α-lattice cell and the tcalculated XRD pattern have been included in the Supporting Information (Fig. S3): (100) at $2\theta$ = 17.9º; (020) at $2\theta$ = 18.3º; (110) at $2\theta$ = 20.0º; and (021) at $2\theta$ = 26.6º [33]. This indicates that the crystalline moieties in these samples are isotropically oriented. The pattern obtained for the PVDF nanotubes also contains only features for α-PVDF, however, in contrast to the bulk samples, the nanotubes display only reflections from the (020) and (110) lattice planes, with the intensity of the (020) reflection being significantly higher than that of (110). Clearly, our PVDF nanotubes feature a strong texture.

Further information on this texture was obtained from XRD pole figure measurements for the (020), (110) and (021) reflections (Fig. 2e). In these, the $\psi$-angle, defined as the angle produced by the rotation of the sample around an axis parallel to the plane of the template surface, is plotted along the radial direction, and the $\phi$-angle, defined as the angle produced by the rotation of the sample around an axis parallel to the long axis of the nanotubes, is plotted along the azimuthal direction (see inset in Fig. 2d). We find the (020) reflection to be centered at $\psi$ = 0º, indicating a preferred orientation of the corresponding lattice planes along the direction perpendicular to the longitudinal axis of the nanotubes, in accordance with the findings by Steinhart et al.[22]. The (110) planes are detected in two different orientations. A fraction of (110) planes lay perpendicular to nanotube long axis. Another fraction of (110) planes are tilted by approximatlye $\psi$ ~ 60º with respect to the longitudinal axis of the nanotubes. Note here that each circular line along the radial direction in the pole figures, i.e. $\psi$- angle, corresponds to 10º. Finally, we observe that most of (021) planes are detected at an angle $\psi$ ~40º with respect to the longitudinal axis of the nanotubes.

Combining all our observations, these data indicate that our nanotubes are mainly comprised of crystalline moieties in which the crystallographic <020> direction aligns with the nanotube long axis, as already alluded to above. These crstyalline lamellae are isotropically oriented along the azimuthal $\phi$-angle, as deduced from the ring-like patterns of (110) and (210) reflections. However, crystals with <*hkl*> directions with a zero *l*-index are only detected along the nanotube long axis direction. In these crystals, the polymer chain lays approximately perpendicular to longitudinal direction of nanotubes, therefore lamellar crystals can grow along the longitudinal and the azimuthal directions of the nanotube; thereby yielding large crystals that can be easily detected in the WAXS experiment, as discussed already by Steinhart et al [22]. As a matter of fact, almost uniaxial texture is observed in our PVDF nanotubes (<020> crystallographic direction aligned with the pore longitudinal axis). Besides the argument above, typically, the formation of these ultra-oriented structures seems to be associated with a low nucleation rate. Thus, under low nucleation rate conditions, such as that used in our experiment, i.e. 1ºC/min, the crystallographic direction with the fastest growth rate usually aligns with the longitudinal axis of the pores. As proposed by Huber [34-36], this can be traced to a crystallization mechanism first suggested by Bridgman for the single crystal growth in narrow capillaries [37], where the crystalline direction with higher growth rate propagate along the direction of the long axis, while other growth directions die out.

**Relaxation processess of nanotubular PVDF**

Having obtained a detailed picture of the microstructure of our PVDF nanotubes, we were able to assess the segmental dynamics of these systems, and use our structural picture to establish structure/property interrelationship and identify the effect that nanoconfinement has on these dynamic processes. In order to maximize the dielectric signal of the segmental relaxation in our PVDF nanostructures, which is due to the amorphous fraction in the sample, we produced nanotubes via rapid cooling from the infiltration temperature to room temperature. This procedure yielded nanotubes with a lower crystallinity and smaller crystals than slowly crystallized nanotubes (Supporting Information Fig S4), and the crystalline moieties were more isotopically oriented in such structures, although the tendency of the system to develop crystals oriented with the <020> direction parallel to the long axis of the nanopores can be still observed.

DS measurements were performed at selected frequencies and temperatures on bulk PVDF and PVDF nanotubes/AAO template assemblies. The dielectric signal recorded of the nanotubes displays, thus, a contribution from both the polymer as well as the template. However, the contribution from the template is two orders of magnitude weaker than that of the PVDF and shows no features (see Supporting Information Fig. S5). Therefore, the signal recorded in our experiments can be interpreted as originated by the molecular relaxation processes in the polymer nanostructure.

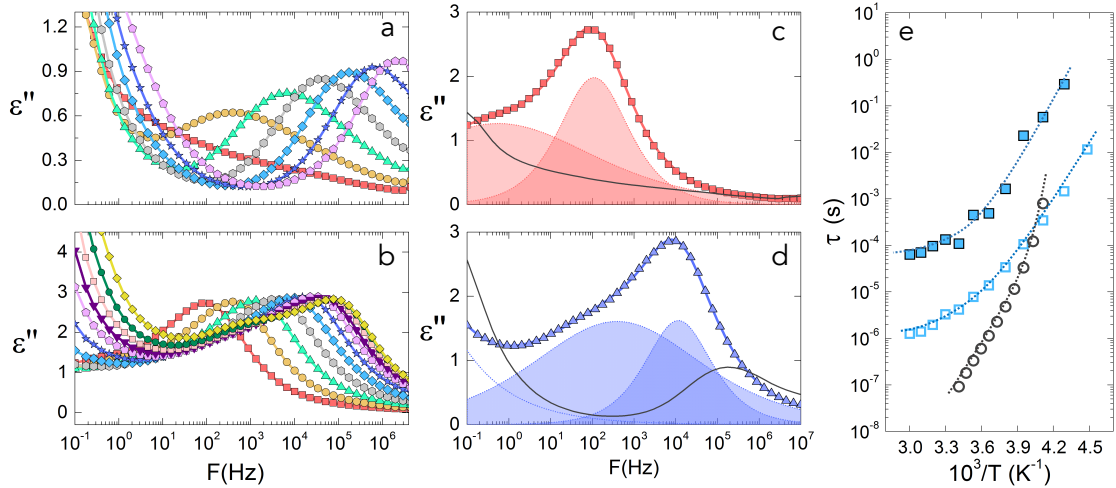

**Figure 3**: (a) ε'' values as a function of frequency for bulk PVDF and (b) PVDF nanotubes, at selected temperatures: -40 ºC (red squares), -30 ºC (orange circles, -20 ºC (green triangles),–10 ºC (grey hexagons), 0 ºC (blue diamonds) 10ºC ,(purple stars), 20 ºC (light purple pentagons), 30 ºC (violet triangles), 40 ºC (pink squares), 50 ºC (dark green circles), and 60 ºC (yellow diamonds). (c) ε'' values as a function of frequency for PVDF nanotubes at T=-40ºC and (d) T=0ºC. The dotted curves indicate separate contributions from the regular bulk-like α and a slower relaxation process, i.e., the interfacial relaxation, which are obtained by fitting two HN equations to the experimental data. For 0 ºC data, an additional conductivity-like contribution has also been considered. Solid black lines correspond to the relaxation curves of bulk PVDF at those temperatures. (e) Relaxation times as a function of the reciprocal temperature for bulk PVDF (grey circles) and PVDF nanotubes (blue squares). Open symbols correspond to the bulk-like α relaxation observed in the nanotubes, whereas full symbols correspond to the relaxation times of the interfacial relaxation. Dotted lines are guides to the eye.

Fig. 3 shows the dielectric loss spectra of bulk PVDF (Fig. 3a) and PVDF nanotubes (Fig. 3b). In the frequency window and the temperature range analyzed (-40 to +20 ºC), the dielectric spectra of the bulk sample are dominated by the α-relaxation process, i.e. the segmental relaxation which is related to the cooperative segmental motions appearing in a supercooled liquid at temperatures above $T_g$. The segmental relaxation is a feature of the non-crystalline parts of the material, which, in the case of typical semicrystalline structures such as bulk PVDF, is sandwiched between lamellar crystals, as stated above.

The dielectric response of the nanotubes shows also a maximum in the frequency range analyzed (Fig. 3b); however, clear differences can be observed when the behavior of both materials are compared. First of all, the positions of the relaxation maxima for the nanotubes seem to present a weaker dependence on temperature than those of the bulk sample. The second difference relates to the shape of the observed relaxation curve. In order to analyse the dielectric signal measured for the nanotubes, the spectra at each temperature were decomposed into different contributions using the Havriliak-Negami formalism discussed in Ref. [38]. Interestingly, we find that unlike the bulk, the dielectric behavior of the PVDF nanotubes can not be fitted with a single relaxation process but rather two processes. This is illustrated in Fig. 3c and 3d, where the two contributions towards $\varepsilon''$, measured for the nanotubes as function of frequency at $T$ = -40 ºC (Fig 3c) and T = 0 ºC (Fig 3d), are given. For the sake of comparison, the relaxation curves for the bulk PVDF at the same temperatures are also shown as solid black lines. The Havriliak-Negami formalism employed provides the following expression for deducing the complex dielectric permittivity of a system:

$$\varepsilon^* = \varepsilon_\infty + \frac{(\varepsilon_0 - \varepsilon_\infty)}{[1+(i\omega\tau_{HN})^b]^c} \qquad (1)$$

where $\varepsilon_0$ and $\varepsilon_\infty$ are the relaxed and unrelaxed dielectric constants, respectively; $\tau_{HN}$ is the characteristic relaxation time, and $b$ and $c$ are parameters which describe the symmetrical and asymmetrical broadening of the relaxation function, respectively [39]. Assuming an underlying distribution of characteristic times for the origin of the broadened relaxation function, the mean relaxation time of the relaxation times distribution function, $\tau$, is calculated using:

$$\tau = \tau_{HN} \left[\sin\left(\frac{b\pi}{2+2c}\right)\right]^{-1/b} \left[\sin\left(\frac{bc\pi}{2+2c}\right)\right]^{1/b} \qquad (2)$$

Fig. 3e shows the relaxation map for bulk PVDF and PVDF nanotubes, where $\tau$ the calculated using Eq 2 are plotted in a logarithmic scale versus the reciprocal temperature. Both the fastest relaxation in the nanotubes well as that of bulk PVDF exhibit a $\tau$ that follows a non-Arrhenius behavior that is characteristic of segmental motions in glass forming systems. This observation, together with the proximity of the characteristic times of these two processes, suggested the assignment os the same origin to both relaxations, i.e., they would correspond to the segmental relaxation of PVDF chains in the amorphous regions sandwiched between crystalline lamellae. Note, though, that this process exhibits a weaker dependence with temperature in the nanotubes compared to that of bulk PVDF indicating a lower activation energy for this process in the nanotubular system. This causes the dynamical process in the nanotubes to become faster than in the bulk at temperatures below -25ºC. A similar tendency was observed by Duran et al. for nanorods of polypeptides within alumina templates [40] as well as in PVDF nanorods [12].

The observed enhancement of the dynamics in nanotubes structures at low temperautres may have different origins; most likely, it can be associated with the lower crystallinity of PVDF nanotubes compared to bulk material. Indeed, it is well known that the segmental dynamics of semicrystalline polymers is highly affected by the presence of crystalline domains next to amorphous regions [41-43]. Crystalline fractions lead to an increase of the relaxation time and a broadening of the relaxation time distribution function [38]. Hence, a lower crystallinity, as observed for the PVDF nanotubes, would result in a decrease of $\tau$. This effect should be even more pronounced when the material is rapidly solidified, as we did for the sample preparation for the DS measurements, as this will decrease the degree of crystallinity further as our X-ray datain the Supporting Information (Fig. S4) illustrates. Furthermore, the presence of an air-polymer interface at the inner side of the nanotubes' walls may assist to accelerate the relaxation process. Such an enhancement of the dynamics at free interfaces is a well-documented phenomenon in the literature [44-48].

The second relaxation process that we detect in the nanotubes is significantly slower than the bulk $\alpha$-relaxation at all temperatures (Figure 3e). A potential origin of this behavior are favorable interactions of some of the amorphous PVDF fractions in the nanotube structres with the alumina interfaces, which lead to a layer at the interface with nanopore walls where the PVDF macromolecules are of restricted mobility. In fact, considering only the negative Hamaker constant of the nanotube/AAO template system (Supporting Information), it is evident that the alumina surface can act as an attractive interface for the PVDF molecules. Moreover, hydrogen bonding between the hydroxylated pore walls and PVDF chains can likely form [49]. Such attractive forces at the very interface with AAO pore walls, would result in the individual PVDF chains to anisotropically coil. This could lead to more efficient packing [50] and, in turn, to a reduction of the molecular mobility of the chain segments, as has been frequently found for polymers and small molecules confined in pores [30, 51-53].

Combining our observations leads to a dynamical picture where two populations of amorphous regions are present in rapidly solidified PVDF nanotubes. The first population exhibits a relaxation process that is similar to that of the bulk material and, thus, can be assigned to the molecular dynamics of interlamellar amorphous regions. The second population seems to result from regions where the macromolecules feature a reduced mobility and, hence, likely are located at the interface with the AAO pore wall as discussed above. We do not detect this population in the slowly crystallized nanotubes; however, we note that such a conclusion is in agreement with the fact that rapid solidification leads to nanotubes of a lower degree of crystallinity (and, hence, possibly two types of amorphous regions) than when they were cooled from the melt with 1 °C/min. It is also in accordance with the relaxation behavior observed in non-hollow PVDF nanorods, which are also of a low degree of crystallinity and display a pronounced bimodal relaxation dynamics [12]. Our picture also agrees with work by Li et al., who reported a double glass transition, $T_g$, for PMMA confined in AAO nanopores of 80 nm in diameter [54]. These authors proposed that PMMA macromolecules near the AAO pore walls display a higher $T_g$ due to strong interfacial interactions, while molecules located in the centre of the nanostructures would display a reduced packing density, leading to a lower $T_g$. Other authors have observed that the contributions of the two different dynamical processes in the same polymer depend on pore diameter [55-56]. This suggests that the interface-to-volume ratio of such nanostructures dictates which dynamical process (in intercrystalline amorphous phase vs. interface to tube walls) dominates over the other. The slower relaxation could be also attributed to a Maxwell Wagner Sillars effect, that appears in inhomogeneous systems and it is associated with a mesoscopic charge separation.[57] All of the above scenarios implies a layer of different mobility attached to the alumina walls.

The results on the confined system in the high-temperature region (above 20 ºC) of the relaxation map (Fig.3e) shows, moreover, an even more striking effect. Fig. 3e shows that as temperature increases, both the bulk-like and the interfacial relaxations tend to become temperature independent. This behavior is very unusual for polymers and has been observed before in relaxor ferroelectric solid solutions of the type $PbMg_{1/3}Nb_{2/3}O_3$-

PbSc$_{1/2}$Nb$_{1/2}$O$_3$-PbZn$_{1/3}$Nb$_{2/3}$O$_3$ (PZN-PMN-PSN) [58] and in quasi one-dimensional ferroelectric systems [59], where it has been explained as resulting from the coexistence of local polar structures in a non polar matrix. [60]

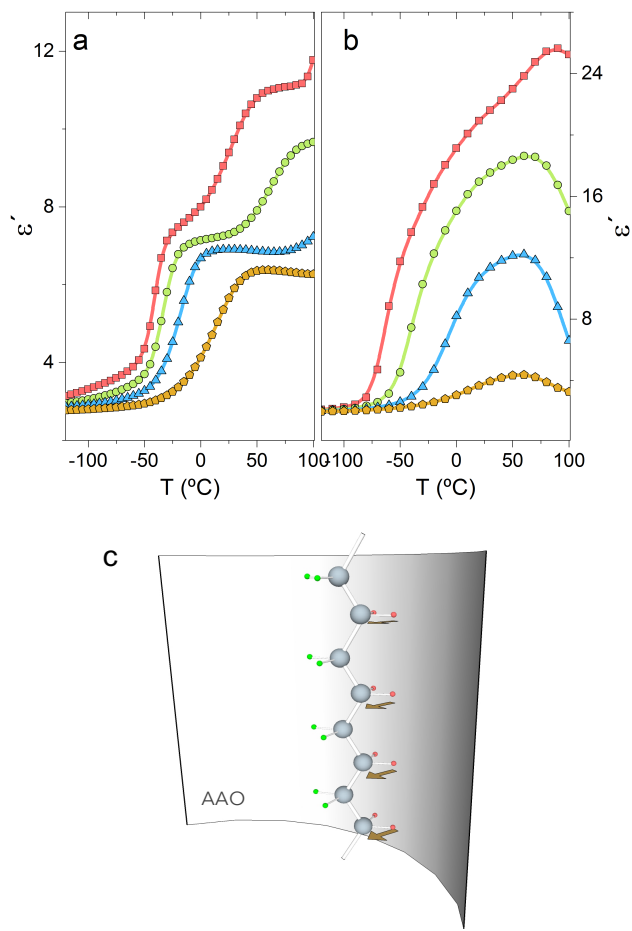

**Fig. 4**. Temperature dependence of the dielectric permittivity ($\varepsilon'$) of bulk PVDF (a) and PVDF nanotubes (b) measured at different frequencies: $10^0$ Hz (red squares), $10^2$ Hz (green circles), $10^4$ Hz (blue triangles), $10^6$ Hz (orange pentagons). (c) Schematic illustration of the conformation a PVDF chain segment might adopt at the interface with the AAO pore wall. Hydrogen atoms face preferentially to the pore walls, while fluorine atoms point towards the centre of the pores (green spheres: fluorine atoms; grey spheres: carbon atoms; red spheres: hydrogen atoms). Hence dipoles in PVDF will be preferentially oriented edge-on with respect to the pore walls (brown arrows), which would lead to have PVDF chains with an overall non-zero dipole moment.

These strong nanoscale interfacial effects that are found in PVDF/AAO template assemblies also have a striking impact on other properties such as the real part of the dielectric permittivity ($\varepsilon'$). Figure 4 shows the behavior of $\varepsilon'$ with temperature for bulk PVDF (Fig. 4a) and PVDF nanotubes (Fig. 4b). The spectrum of the bulk material is characterized by an intial step-like increase of $\varepsilon'$ at temperatures around -40 ºC. This feature can be attributed to the segmental relaxation of PVDF. A second increase of $\varepsilon'$ occurs at higher temperatures, which results from the activation of local motions in the crystalline phase, i.e. the $\alpha_c$ relaxation, known to occur in the $\alpha$-PVDF [61]. The behavior of $\varepsilon'$ with temperature is strikingly different for the PVDF nanotubes. While the step-like increase of $\varepsilon'$ in the low temperature regime is similar to that of the bulk (although it occurs at notably lower temperatures: i.e. -70 ºC compared to -40 ºC observed for bulk PVDF for F ~10 Hz), the high-temperature-region of the nanotube spectrum exhibits a broad maximum. Interestingly, in line with the behavior of $\varepsilon''$ in the high temperature region, the behavior of $\varepsilon'$ also resembles that of relaxor ferroelectrics [62-63]. Relaxor ferroelectrics are ferroelectrics with a broadened phase transition that it extent over a given range of temperature. They normally exhibit high values of its dielectric constant, and they are strongly linked to disorder. The permittivity of relaxors exhibit a maximum that shifts with temperature. However, our PVDF nanotubes seem to crystallize into the $\alpha$-phase, which is well-known to be paraelectric (Supporting Information Fig. S4) and, thus, should not display a ferro-para transition.

A comprehensive study focused on this striking ferroelectric-like behavior is currently ongoing. Here we advance some hypotheses that can be to considered in order to rationalize our observations:

1). A decrease of $\varepsilon'$ in the high-temperature-region of the dielectric spectra can principally result from the progressive crystallization upon heating of parts of the amorphous fraction in PVDF nanotube structures. However, the reduction of the amount of mobile dipoles as a consequence of crystallization would be accompanied by a decrease of the $\varepsilon''$ intensity – a feature we do not observe in our data (Supporting Information Fig. S6), rendering this potential origin of our observations unlikely.

2). The decrease in $\varepsilon'$ may also result from a ferro-to-paraelectric transition within initially polar (ferroelectric) crystalline regions that potentially could form at the very interface with the AAO pore walls. Potentially, strong dipolar interactions between PVDF macromolecules and the hydroxylated AAO pore walls might promote the nucleation of a polar phase (*β*, *γ* or *δ*) under rapid crystallization conditions, in which surface nucleation processes may be relevant. Examples where interfacial phenomena provoke the crystallization of polar PVDF phases have been described in the literature [64-65]. Our WAXS data imply, however, that the nanotubes are essentially comprised of *α*-PVDF, which is a non-polar phase (Suporting Inormation Fig. S4); although we can not exclude that the palar *δ*-phase is formed [66-67]. Unfortunately, the signal-to-noise level of the WAXS patterns did not allow us to perform a detailed phase analysis. We like to note, though, that we do not observe features that could be assigned to a Curie transition that should occur at temperatures around 50-70 ºC in case a polar phase is present, ruling essentially this explanation out as well.

3). Another, plausible explanation for the observed decay of *ε´* in the high temperature range of the spectra requires consideration of the chain conformation of amorphous PVDF macromolecules close to the AAO pore walls. As mentioned above in our discussions of the dynamic behavior of PVDF nanotubes, both dispersive and polar attractive interactions are expected to occur between a molecule with strong dipoles like PVDF and a protic polar material, such as the hydroxylated pore walls of AAO templates. These include from a general Lewis perspective acid-base interactions, i.e. lone pair donor/acceptor interactions, as depicted in Fig. 4c. As such, two kinds of hydrogen bonds are subject to be formed between PVDF and the AAO: $_{(polym)}$C–H •••• O- $_{(AAO)}$ ; and $_{(polym)}$C–F •••• H-O- $_{(AAO)}$. However, wettability studies of fluorinated polymers on polar protic liquids have suggested that the dominant hydrogen bonds are of the type $_{(polym)}$C–H •••• O- $_{(AAO)}$; in fact, the formation of other types of hydrogen bonds has been questioned [49].

As a consequence of these attractive forces, regions can develop within the nanotubes where the PVDF macromolecules will form anisotropical coils with a somewhat longer axis parallel to the AAO surface [50]. Moreover, due to the specificity and the high directionality of the $_(hydrogens bonds ( of the type $_{polym)}$C–H •••• O- $_{(AAO)}$ ) that form between the two components, a major population of PVDF macromolecules will adopt a chain conformation where the hydrogen atoms face preferentially towards the pore walls, while the fluorine atoms will point towards the opposite direction, as schematically illustrated in Fig. 4c and similar to conformations that have been proposed for PVDF molecules close to other polar species such as water [65]. In such an arrangement, the

dipoles of the PVDF chains close to the AAO pore wall would be preferentially oriented edge-on with respect to the pore walls producing PVDF regions with a non-zero dipole moment at the interface with the AAO. In this regions, the chains must, however, be mobile – at least at the length scale of the segmental relaxation and within the temperature regime where we observed the notable decrease in $\varepsilon'$. At a larger length scale, the hydrogen bond network might induce an "anisotropic liquid"-like behavior that could lead to local ferroelectric domains. As a matter of fact, the decrease in $\varepsilon'$ may be explained in terms of a loss in anisotropy in the PVDF chain conformation in these regions upon exposure to more elevated temperatures as this would reduce the average number and of the H-bonds between the macromolecules and the AAO walls as well as their strength. In such a scenario $\varepsilon'$ would depend on the attachment/detachment dynamics of the H-bond structure, which agrees with our observation that the maximum of $\varepsilon'$ is frequency-dependent in our PVDF nanotubes. Interestingly, such a frequency dependence of the position of the $\varepsilon'$ maxima is a signature of relaxor ferroelectrics. Also, in line with this explanation, as mentioned before, around 60 ºC, the relaxation starts to slow down as temperature increases. Similar behavior has been found for the ferroelectric charge transfer salt $(TMTTF)_2AsF_6$,[59, 68] which is a quasi one-dimensional organic conductor with a ferroelectric transition. The authors found a critical increase of the mean relaxation rate as temperature increased. They explained this critical behavior as derived from the damping of a soft mode.[59, 69]

These are technologically attractive because they feature a broad temperature range where the ferro-to-para transition occurs leading to a small hysteresis for this transition[70] in contrast to 'normal' ferroelectrics which can display a pronounced hysteresis. This behavior clearly can be manipulated: here we induced it by rapid solidification of PVDF in AAO templates to reduce the amount of crystalline fraction that is formed. Likely, it can also be controlled by changing the confinement conditions. Indeed, an even more pronounced frequency dispersion was observed in 60-nm-diameter PVDF nanorods[71] where confinement effects might be more pronounced than in our nanotube structures.

**Conclusions**

We have elucidated here the impact of the tubular confinement on the structure and dynamics of the semicrystalline polymer PVDF. Our microstructural analysis suggests that the slowly crystallized PVDF nanotubes are comprised of alternating crystalline lamellae and amorphous regions commonly found in semicrystalline polymers. The periodicity of the multilayer structure is slightly larger than that of the bulk material, as the stack must accommodate a larger amount of amorphous PVDF chains segments. Confinenent leads to a strong texture with the crystalline <020> directions being aligned with the long axis of the.

Most importantly, the dynamical picture of rapidly crystallized nanotubes reveals the presence of two populations of amorphous PVDF fractions. In addition to a phase that relaxed similarly to bulk PVDF, a significant amorphous chain population exists at the interface with the AAO pore wall that, due to the attractive interaction with the pore walls, has a significantly slower relaxation dynamics than the bulk material. Strickingly, both the bulk-like and the interfacial relaxation tend to become temperature independent as the temperature increases; a behavior that is very unusual for polymers and has been observed before in inorganic relaxor ferroelectrics. In line with this, the real part of the dielectric permittivity exhibits a broad maximum when plotted agains that temperature, that also resembles that of relaxor ferroelectrics. Therefore, we suggest that ferroelectric domains are forming at the interface with the AAO pore walls. They result from hydrogen bond formation that induces an anisotropic chain conformation and a preferred orientation of local dipoles in the PVDF macromolecules in these regions. This arrangement seems to be the origin of the relaxor-ferroelectric-like behavior of PVDF nanotubes as a direct consequence of confinement and rapid solidification both leading to a lower degree of crystallinity and the former providing the critical interfaces for nanoscale interactions that can induce chain conformation changes. This also means that this behavior can be manipulated by selection of AAO templates of different pore dimensions (and, hence, the interface/volume ratio) and solidification conditions selected. Clearly, this relaxor-ferroelectric-like response which has not been observed so far for the neat PVDF homopolymer, may open up a series of new possibilities for the usage of this 'plastic', for example the fabrication of electrocaloric devices for solid-state refrigeration where a relaxor ferroelectric behavior is desirable.

**Associated Content**

*Supporting Information*

The Supporting Information is available free of charge on the ACS Publications website

Calculation of the mass of PVDF contained inside the nanopores, thermogravimetric curves of PVDF, lattice cell of α-PVDF, analysis of the Hamaker constant for the substrate/polymer/air system, WAXS pattern of the rapidly solidified samples, dielectric response of the AAO templates, further ε" vs F plots.

**Acknowledgements**

J. M. acknowledges support from the European Union's Horizon 2020 research and innovation programme under the Marie Skłodowska-Curie grant, agreement No 654682. A. A. and A. I. acknowledge financial support from the Spanish Ministry 'Ministerio de Economia y Competitividad (MINECO), code: MAT2015-63704-P (MINECO/FEDER, UE) and by the Eusko Jaurlaritza (Basque Government), code: IT-654-13. A. C would like to acknowledge the financial support of the EPSRC (EP/M014142/1). A. N, T. E. and C. M acknowledge financial support form MINECO (codes: MAT2014-59187-R, MAT2015-66443-C02-1-R MAT 2014-53437-C2-1P, respectively). N. S. is in addition grateful for support by a European Research Council ERC Starting Independent Research Fellowship under the grant agreement No. 279587.

**TOG Graphic**

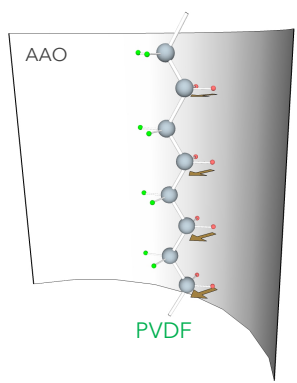